\documentclass[%
 aip,
 amsmath,amssymb,
 reprint,%
]{revtex4-1}

\usepackage{bm}
\usepackage{braket}
\usepackage{graphicx}
\usepackage{dcolumn}% Align table columns on decimal point

\usepackage{xcolor}

\usepackage[utf8]{inputenc}
\usepackage[T1]{fontenc}
\usepackage{mathptmx}
\DeclareUnicodeCharacter{2212}{-}
\begin{document}

\preprint{AIP/123-QED}

\title[Dissociation in a strong field: a quantum analysis of the relations between angular momentum and angular distribution of fragments.]{Dissociation in strong field: \\a quantum analysis of the relation between angular momentum and angular distribution of fragments.}
% Force line breaks with \\

\author{Bar Ezra}
 \email{bar.ezra@mail.huji.ac.il}
\affiliation{The Institute of Chemistry and the Fritz Haber Centre for Theoretical Chemistry, The Hebrew University of Jerusalem, Jerusalem 9190401, Israel}%

\author{Shimshon Kallush}%
\email{shimshonkallush@braude.ac.il}
\affiliation{The Institute of Chemistry and the Fritz Haber Centre for Theoretical Chemistry, The Hebrew University of Jerusalem, Jerusalem 9190401, Israel}%
\affiliation{Department of Physics and Optical Engineering, ORT-Braude College, P.O. Box 78, 21982 Karmiel, Israel }

\author{Ronnie Kosloff}
\homepage{https://scholars.huji.ac.il/ronniekosloff}
\email{ronnie@fh.huji.ac.il} 
\affiliation{The Institute of Chemistry and the Fritz Haber Centre for Theoretical Chemistry, The Hebrew University of Jerusalem, Jerusalem 9190401, Israel}%

\date{\today}

\begin{abstract}
An ab initio simulation of strong-field photodissociation of diatomic molecules was developed, inspired by recent dissociation experiments of \(F_2^-\). 
The transition between electronic states was modeled, including the laser pulse and transition dipole, and the angle between them. The initial conditions of the system were set to be thermal and to include different rovibrational states. Carefully designed absorbing boundary conditions were applied to describe the boundary conditions of the experiment.
We studied the influence of field intensity on the direction of the outcoming fragments and laboratory-fixed axis, defined by the field polarization. 
At high intensities, the angular distribution became more peaked with a marginal influence on kinetic energy release.
\end{abstract}
\maketitle

\section{Introduction}

High-intensity time-domain spectroscopy can be modeled by explicitly solving the time-dependent Schr\"{o}dinger equation, for which knowledge of the full Hamiltonian of the process is essential. Employing the Born--Oppenheimer approximation, the Hamiltonian is separated into electronic and nuclear terms \cite{born1927quantentheorie}. The nuclear degrees of freedom are represented by vibrational and rotational terms. Coupling the external electromagnetic field and the transition dipole moment, induces transitions between electronic states. Each electronic transition is accompanied by an angular momentum change of one unit of \(\hbar\). \par

In this paper, we study strong-field spectroscopy of diatomic molecules \cite{suits2018invited}. 
%Specifically, we study the photo dissociation of \(F_2^-\), which might be considered a case study representing other diatomic ions.
The most studied molecule of these molecules is the diatomic anion  $I_2^-$, which has been subjected to ultrafast photoelectron spectroscopy. Models describing its photoelectron distribution have been constructed under different approximations \cite{Zanni1999femtosecondI2}. It is quite customary to neglect the rotational dependency of \(I_2^-\), due to iodine’s significantly heavy nuclei and slow dynamics within the \(fsec\) pulse time scales. \par

A semi-analytical theory of photodissociation processes was first developed by Zare \cite{zare1972photoejection}, and then further discussed and extended \cite{choi1986theoryBernstein,seideman1996analysisSeideman}. The semi-analytical assumptions include only one photon transitions. Hence, the results were suited to low-intensity radiation, where only single-photon processes are considered to lead to the dissociation of the molecule. \par

An extended theoretical treatment was developed by Ben-Itzhak, with Esry's collaboration, employing ab initio simulations to interpret the experimental results of $H_2^+$ photodissociation \cite{wang2006dissociationH2, anis2008roleH2, mckenna2012controllingH2}. Initially, they considered only electronic and vibrational modes, assuming that the influence of the nuclear rotational states was negligible \cite{wang2006dissociationH2}. This assumption was revisited by Anis and Esry in a later study \cite{anis2008roleH2}, where they concluded that the rotational states are essential. Note, that the initial state was set as the ground rovibrational state, practically setting the temperature to zero, an assumption that is justified for such species at low temperatures. Furthermore, they employed the Floquet dressed-states picture which is appropriate when the laser field is characterized by a single frequency. The general issue of diatomic dissociation is currently active applying different approximations \cite{csehi2019ultrafast,sun2019mappingK2,mcdonald2016photodissociation,halasz2015direct}.

In this paper, we had two main goals: first, to acquire a more fundamental comprehensive understanding of the high-field dissociation process on a quantum scale; and second, to simulate the angular distribution of the photo-fragments resulting from this process. The current simulation uses the physical parameters of $F_2^-$. The motivation to use the $F_2^-$ molecule as a benchmark follows from a recent experimental study by Strasser \emph{et al.}\cite{shahi2017intense,shahi2019ultrafast}. Furthermore, its electronic structure is rather simple: there are only four low-lying states before reaching the detachment continuum. The photoelectron detachment channel is thus ignored, due to the $10\,eV$ energy gap to the closest ionic state.

\begin{figure}
\includegraphics[width=80mm]{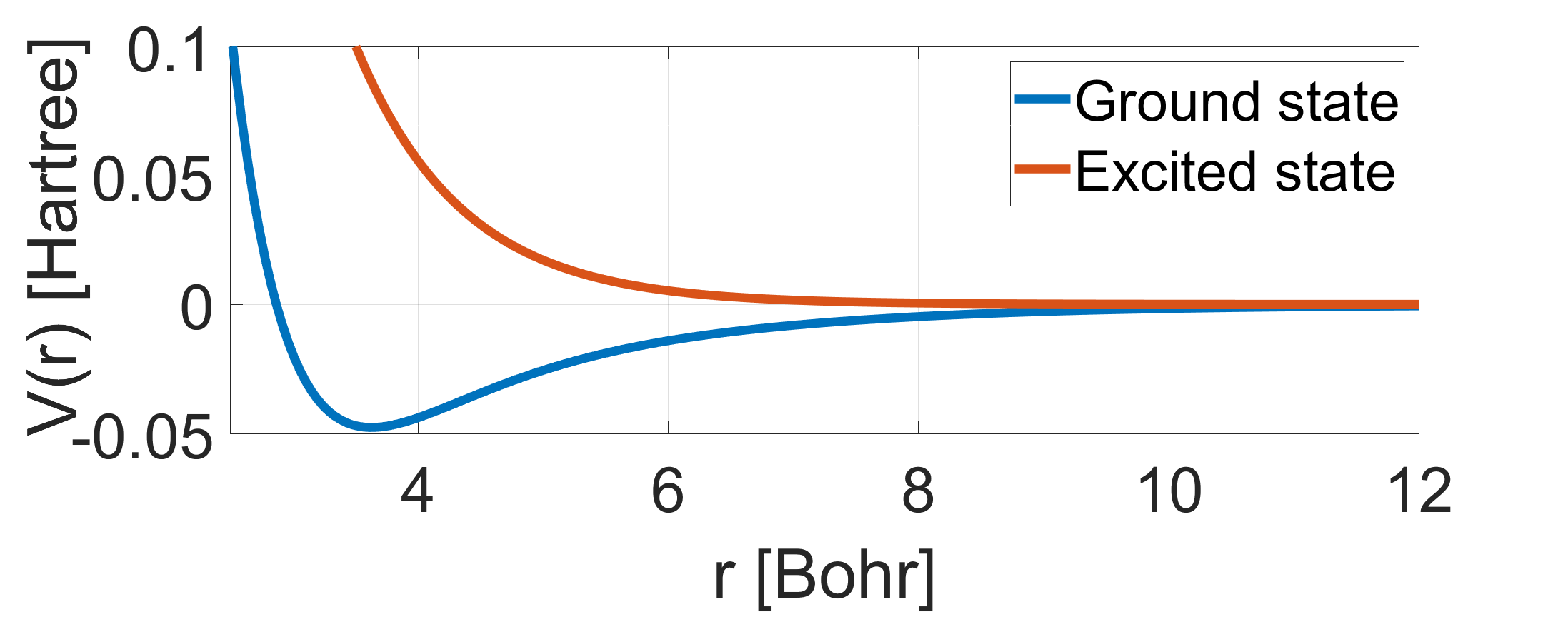}
    \caption{Ground and excited electronic state potentials used in the calculation. The ground state, blue, is chosen as $\Sigma$ symmetry. The excite state, red, is chosen as either $\Sigma$ or $\Pi$. These potentials represent the computed $\sigma_u$ and $\sigma_g$ state taken from \cite{Moszynski}}
    \label{fig:potential}
\end{figure}

We present a first principle model, which includes a numerically exact solution of the time-dependent Schr\"{o}dinger equation. All of the nuclear rotational and vibrational states, and the couplings between them, are included. We simulate a finite temperature ensemble, and therefore the initial state is thermal. 
The option of the Floquet dressed picture is not employed of the reason that it is limited to relative long pules. 
%The Floquet dressed-states picture is not used here, for the purpose of generalizing the model. 
The angular distribution of the fragments, as well as other observables, are extracted directly from the obtained wavefunction. We used the freedom of setting the model to explore fundamental processes and simplify the \(F_2^-\) system into a two-electronic-state system, presented in Fig.\ref{fig:potential}. The considered model simplifies the full physical model, to enable clear and informative bench-marking before going down the rabbit hole of simulating a full-dissociation process.

\section{Model}
\begin{figure}
\includegraphics[width=50mm]{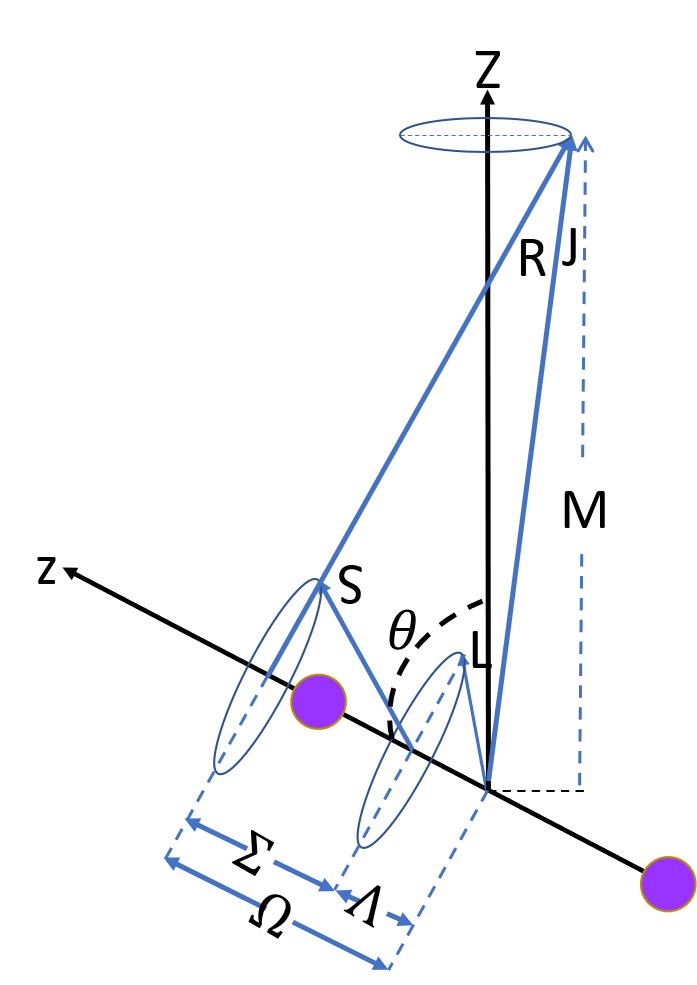}
    \caption{Hund's case (a). The angular momentum components on a diatomic framework. The inter-molecule axis are marked as \(z\) and the polarization of the field axis as \(Z\). The angle between the two axis defined as \(\theta\). Figure \ref{fig:state} presents the full coordinates basis, both inter-nuclear and lab, with all related Euler angles.   }
    \label{fig:quantum_number}
\end{figure}

We consider a model of a diatomic molecule in the gas phase (See Fig. \ref{fig:quantum_number}).
The quantum dynamics takes place on \(n\) electronic states and three ro-vibrational internal nuclear degrees of freedom. The spin was assumed to be zero, and thus does not contribute to the system complexity. 
\par
The Hamiltonian of the system contains the nuclear Hamiltonian of all of the electronic states, and coupling elements between the states due to the transition dipole moment. The Hamiltonian in a Born--Oppenheimer expansion \cite{born1927quantentheorie} of electronic and nuclear coordinates becomes:

\begin{equation}
    \hat{H}_{sys} = \sum_n \left( \hat{H}_n\otimes\left|n\right>\left<n\right| + \sum_{k\neq n} \boldsymbol{\mu}_{n,k}\cdot\boldsymbol{\varepsilon}\otimes\left|n\right>\left<k\right|\right)
    \label{eq:Hamiltonian}
\end{equation}
where \(\hat{H}_n\) is the nuclear Hamiltonian operator of the electronic state \(n\), \(\boldsymbol{ \mu }_{n,k}\) is the transition dipole operator between the states \(n\) and \(k\), and
\(\boldsymbol{\varepsilon}\left(t\right)\) represents the laser field. Note that both \(\boldsymbol{\varepsilon}\) and \(\boldsymbol{\mu}\) are vectors, and the scalar product between them results in different transition schemes, which depend on the involved states' symmetries and the field polarization. In cases where the transition between the states is forbidden, the dipole element vanishes. Non-adiabatic coupling can be added to the equation \ref{eq:Hamiltonian}. \par
The nuclear Hamiltonian of each electronic state can be written as \par

\begin{equation}
    H_{n}=\frac{\hat{P}_{r}}{2m_r}+V_n\left(r\right)+\hat{H}_{rot} \left(r\right)
\end{equation}
where \(r \) is the internuclear distance, \(\hat{P}_r\) is the corresponding momentum, $m_r$ is the nuclear reduced mass, \(V_{n}\left(r\right)\) is the electronic potential of the state \(n\), and \(\hat{H}_{rot}\) is the rotational Hamiltonian. The molecular and linearly polarized laser field axes define the internuclear axis, \(\hat{z}\), and the laboratory axis, \(\hat{Z}\), respectively.\par

The rotational Hamiltonian is
\begin{equation}
    \hat{H}_{rot}\left(r\right)=B\left(r\right)\boldsymbol{\hat{R}}^{2}
\end{equation} 
where \(B\left(r\right)\) is related to the moment of inertia, \(B=\frac{1}{2\mu r^2}=\frac{1}{2I}\) and \(\boldsymbol{\hat{R}}\) is the nuclear rotational angular momentum operator, equal to \(\boldsymbol{\hat{R}=\hat{J}-\hat{L}-\hat{S}}\); \(\boldsymbol{\hat{J}}\) is the total angular momentum of the molecule, \(\boldsymbol{\hat{L}}\) is the electronic orbital angular momentum, and \(\boldsymbol{\hat{S}}\) is the electronic spin angular momentum. \par
In Hund's case (a) \cite{richard1988angular} -- the most common case and the one used here, \(\boldsymbol{\hat{L}}\) and \(\boldsymbol{\hat{S}}\) are coupled to the internuclear axis, \(\hat{z}\), and the projections on it are \({\Lambda}\) and \({\Sigma}\), respectively. The projection of \(\boldsymbol{\hat{J}}\) on the laboratory axis, \(\hat{Z}\), is \({m}\) and on the internuclear axis, \(\Omega = \Lambda + \Sigma\).

\subsection{Basis set representation}
The state of the molecule is described by a nuclear and an electronic state. In the present work, we concentrate on cases in which all electrons are paired, and therefore the spin quantum number is \(\boldsymbol{\hat{S}}=0\). In this situation,  the projection of the electronic orbital angular momentum on the internuclear axis is similar to the total projection, \(\Lambda = \Omega \).\par
\begin{figure}
  \includegraphics[width=50mm]{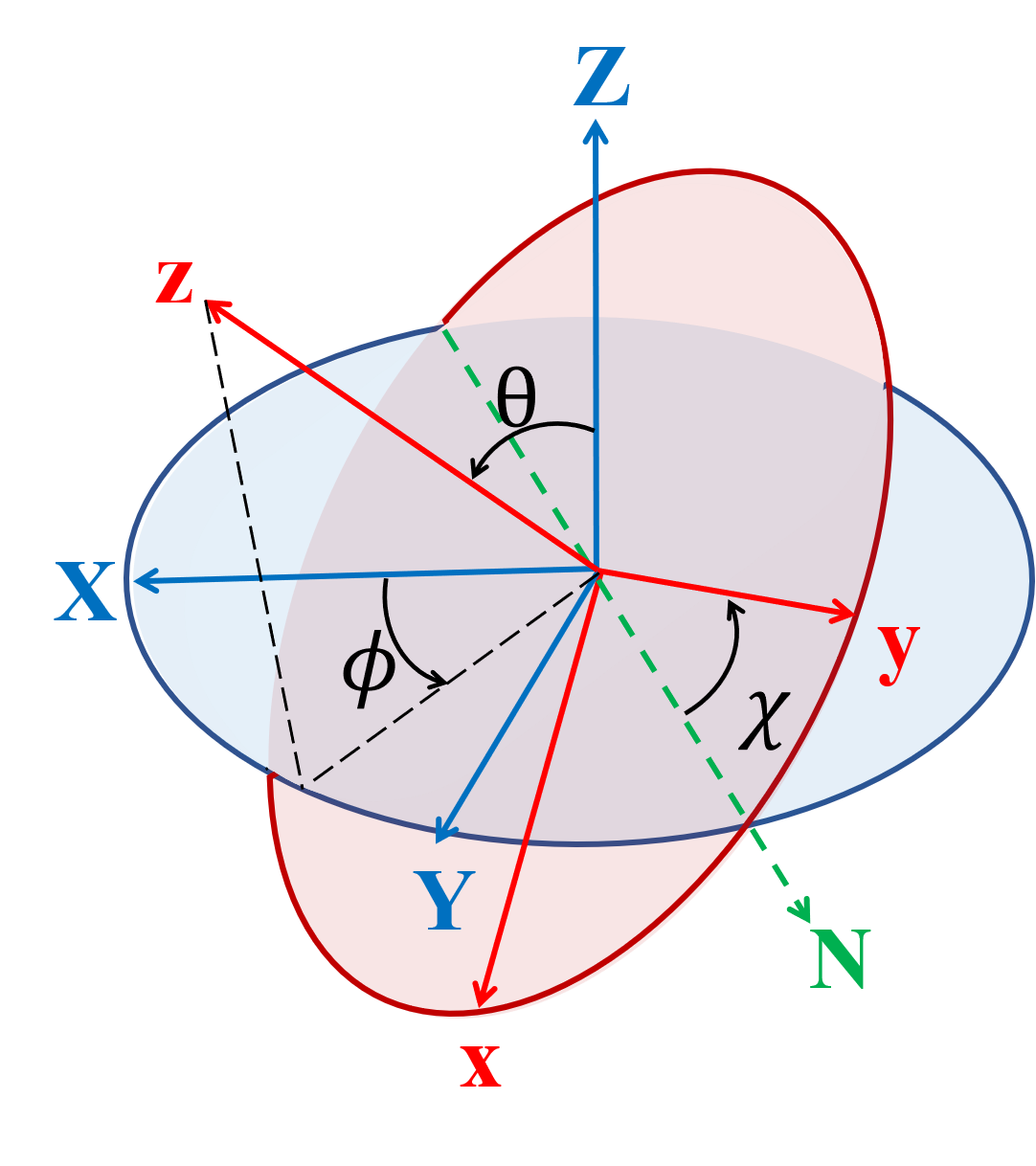}
  \caption{Euler angles relating the space fixed \(XYZ\) and molecule-fixed \(xyz\), for linear molecules the angle \(\chi\) is equal to zero.}
  \label{fig:state}
\end{figure}

The internuclear degree of freedom of the molecule is described by $r$, the distance between the atoms. The orientation is defined by the angles \(\theta\) and \(\phi\) (see Fig.\ref{fig:state}), expanded in terms of the Wigner rotation matrix, \( D^j_{m,\Omega}\left( \phi,\theta \right) \). At \(\Omega =0 \), the matrices are equal to spherical harmonics functions, \(Y_{j,m}\left( \phi,\theta \right) \). In this basis set each state is described by different \(r,j,m\) and \(\Omega\) values. 
The molecular state at time \(t\) can be written by the density matrix: 
\begin{multline}
    \rho\left(r,r',\theta,\theta',\phi,\phi',n,n';t\right)=\\
   \sum_{\zeta,\zeta'}
    \ket{n}\bra{n'} \otimes
    a_{\zeta,\zeta'} \left(r,r';t\right) \otimes \ket{\zeta}\bra{\zeta'}
\end{multline}
where $\ket{\zeta}$ is the wavefunction with the quantum numbers $\ket{j,m,\Omega}$, expanded by the $D^j_{m,\Omega}$ matrices.
\subsection{Coupling elements}

The coupling elements are calculated by vector multiplication between the laser field \(\boldsymbol{\varepsilon}\left(t\right)\) and the transition dipole between the electronic states, \(\boldsymbol{\mu}\). The transition dipole elements depend on the symmetry, \(\Sigma\) or \(\Pi\), and the parity, Gerade or Ungerade, of the states.  \par
For a linearly polarized field, the rotational coupling elements are given by:
\begin{multline}
   \bra{\zeta'} \mu_{0q}\varepsilon_0 \ket{\zeta} \propto\\
   \int D_{m'\Omega'}^{j'\,\,*}\left(\theta,\phi,0\right) \, D_{0q}^{1}\left(\theta,\phi,0\right) \, D_{m\Omega}^{j}\left(\theta,\phi,0\right) \, d\Omega = \\
    = 8\pi \left( \begin{array}{ccc}
        j     & 1     & j'    \\ 
        -m     & 0     & m'
    \end{array} \right)\left( \begin{array}{ccc}
        j     & 1     &j'    \\ 
        -\Omega    & q     & \Omega'
    \end{array} \right)  
\end{multline}
where \(\mu_{0q}\) is the transition dipole moment proportional to \(D_{0q}^{1}\left(\theta,\phi,0\right)\) \cite{band2004rotationalKallush}. The value of \(q\) is determined for a given transition case and is equal to \(q=\Omega-\Omega'\).\par
The first term is the initial state \(D^{j'}_{m',\Omega'}\left(\theta,\phi\right)\), the second term is the coupling operator of  the transition \(D^{1}_{0q}\left(\theta,\phi\right) \), and the third term is the final state \(D^{j}_{m,\Omega} \left(\theta,\phi\right)\). The coupling components are calculated by the given $3-j$ symbols.\par
We will consider two transition cases: \par
\begin{itemize}
\item{ Transition between two \(\Sigma\) electronic surfaces: in this case, \(q = 0-0=0\) hence the coupling \(D\) matrix is \(D_{00}^{1}\). Note that the transition between two \(\Pi\) electronic surfaces gives the same coupling elements, \(q = 1-1=0\).}
\item {Transition between \(\Sigma\)  and \(\Pi\) electronic surfaces: in this case, \(q = 1-0=1\), hence the second \(D\) matrix is \(D_{0\pm{1}}^{1}\).}
\end{itemize}
For the two cases:
\begin{equation}\label{couplingCases}
    \boldsymbol{\mu \cdot \varepsilon } \propto
    \begin{cases} 
        \mu \varepsilon \, D^1_{00}\left(\theta,\phi\right) \propto \, cos\left(\theta\right)    & \Sigma-\Sigma \\ 
        \mu \varepsilon \, D^1_{0\pm1}\left(\theta,\phi\right) \propto \, sin\left(\theta\right) & \Sigma-\Pi
    \end{cases}
\end{equation}
In all the calculations, we assumed that $\mu$ is independent in the internuclear distance $r$.

Additional insight into these results can be obtained by decomposing the transition dipole into a component in the internuclear direction \(\mu_{||}\) and a component in the perpendicular direction \(\mu_{\perp}\). The scalar product between the transition dipole and the field becomes:
\begin{equation}
    {{\mu} \cdot \varepsilon } = \varepsilon\left(\mu_{||}\cos{\theta} +\mu_{\perp} \sin{\theta}\right)
\end{equation}

\begin{figure}
  \includegraphics[width=80mm]{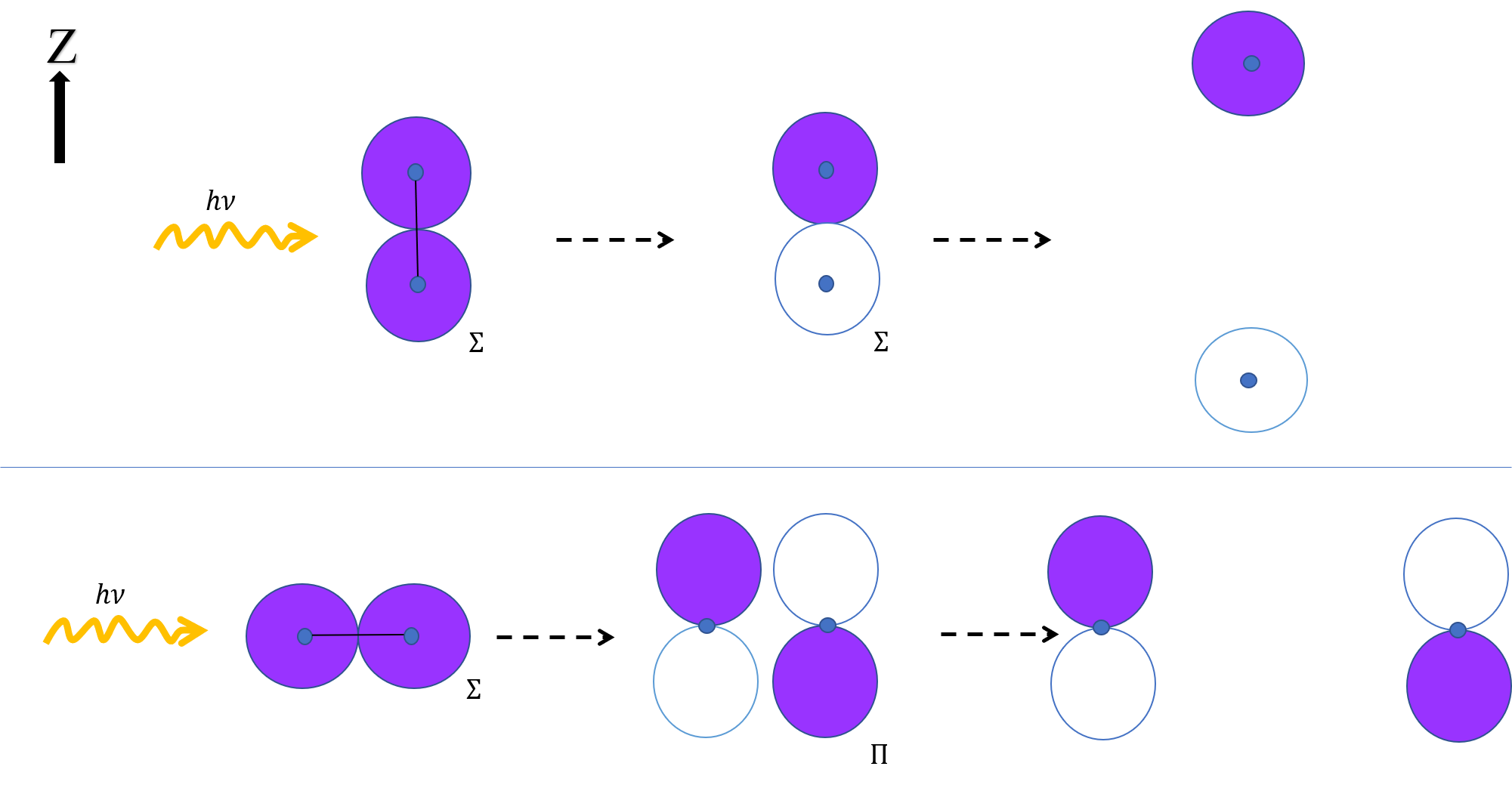}
  \caption{The two transition cases - semi-classical illustration. \textbf{Upper panel}: transition between two \(\Sigma\) electronic surfaces. Starting at a bound electronic surface, \(\Sigma_g\), the laser excites to the un-bounding electronic surface, \(\Sigma_u\), then the molecule dissociated. The fragments are obtained in the direction of the propagation of the light. \textbf{Lower panel}: Transition between \(\Sigma\) and \(\Pi\) electronic surface. Starting at a bound electronic surface, \(\Sigma_g\), the laser excites to the un-bounding electronic surface, \(\Pi_u\), then the molecule dissociated. The fragments are obtained in the perpendicular direction to the propagation of the light.}
  \label{fig:transitionCase}
\end{figure}

Therefore, the transition dipole is parallel in the first case and in the second case to the internuclear direction. An illustration of the two coupling cases is presented in Fig.\ref{fig:transitionCase}.\par
The transition between the electronic states changes the total angular momentum state. For linearly polarized light, the projection of the laboratory axis \(m\) on the $Z$ spatial axis is conserved. The quantum number \(\Omega\) can change with the change in value of \(q\), depending on the transition case. The quantum number \(j\) changes by one unit of angular momentum \(\hbar\) or remains the same (Q branch at the \(\Sigma-\Pi\) transition) . 

\subsection{Initial state}

The model is constructed to mimic a hypothetical experiment: a molecular beam with a fixed temperature is subjected to a pulsed laser source. The ensemble's temperature was taken to correspond to the initial state of \(F_2^-\) at \(20\,K\). At this temperature, the electronic and vibrational degrees of freedom are in their ground state. The initial state can be written as follows:
\begin{multline}
    \rho\left(t=0\right)=\rho_{electronic}\otimes\rho_{rovibration}
\end{multline}
where \(\rho_{electronic}=\left|\psi^e_0\right>\left<\psi^e_0\right|\). The vibrational projection is conditioned on the rotational quantum number $j$, \(\rho_{rovibration}=P_{0,j}(r)=\left|\Phi_{n=0,j_i}\left(r\right)\right>\left<\Phi_{n=0,j_i}\left(r\right)\right|\). The vibrational ground state for each value of \(j\) is obtained by imaginary time propagation  \cite{kosloff1986PropImagTime}, independently for each $j$. Note that the vibrational wavefunction can be different for different quantum numbers.\par

 The initial rovibration state is a Boltzmann distribution of several rotational components.
\begin{multline}
\rho_{rovibration} = \frac{1}{Z}\sum_{\zeta_i} \left(2j_i+1\right)exp\left(-\beta E_{j_i}\right)P_{0,j}(r) \left|\zeta_i\right>\left<\zeta_i\right|\\
      = \sum_{\zeta_i} \xi\left(j_i\right)P_{0,j}(r) \left|\zeta_i\right>\left<\zeta_i\right|
\end{multline}
where $\xi\left(j_i\right)$ is the normalized probability for state $j_i$ with the partition function $Z = \sum_l \left(2l+1\right)exp\left(-\beta(E_l)\right)$. $E\left(j_i\right)$ are the rotational energy eigenvalues, $\hat{H}_{rot} \left|\psi_{\zeta_i}\right> = E \left( j_i \right) \left|\psi_{\zeta_i} \right>$, that depend only on the total angular momentum $j$ as explained above. The index $i$ describes the initial quantum numbers of the state and will be used in this notation from this point on.

%%% להמשיך מכאן
The total initial state is:
\begin{multline}
     \rho\left(r,r,\theta,\theta',\phi,\phi',n,n';t=0\right)=\\ \sum_{\zeta_i} \xi\left(j_i\right) a_{\zeta_i,0,\zeta_i,0}\left(t=0\right) \cdot \left|0\right>\left<0\right|   \otimes \\ 
     P_{0,j}(r) \otimes \left|\zeta_i\right>\left<\zeta_i\right|
\end{multline}
%Where we add the time dependence of \( \left|\Phi\right>\).

\subsection{Propagation in time}

The dynamics of the state is carried by a first principle approach.
The evaluation of the state is described by a unitary transformation:
\begin{equation}
    \rho\left(t\right)=\hat{U}\left(t\right)\hat{\rho}\left(t=0\right)\hat{U}^\dagger\left(t\right)
\end{equation}
where \(\hat{U}\left(t\right) \)  is the propagator which is generated by the time-dependent Schr\"odinger equation: 
\begin{equation}
  i \hbar \frac{d}{dt} \hat{U}\left(t\right) = \hat H(t) \hat{U}\left(t\right)
\end{equation}
and the Hamiltonian is the generator. The density operator, \(\rho(t)\), provides all observable data. For example, the observable result that corresponding to the operator \(\hat{A}\), will be calculated as follows:
 \begin{multline}
     \left<\hat{A}\right>_t= tr\left(\hat{A}\rho\left(t\right)\right) \\ =tr\left(\hat{A}\hat{U}\left(t\right)\rho\left(t=0\right)\hat{U}^\dagger\left(t\right)\right)
 \end{multline}

Under common molecular beam conditions, typical incoherent processes such as collisions or spontaneous decay take place on the time scales of \(t \ge 1\,nsec\). The dissociation process is completed by \(\sim 100\,fsec\). We can therefore assume that the dynamics is fully coherent. The expectation values can be evaluated by using the basis where the initial density operator is diagonalized \(\left|\Tilde{\psi}_{n,\nu,i}\right>=\left|\psi^e_n\right> \otimes \left|\psi^\nu_\nu\right> \otimes \left|\zeta_i\right>\), giving:
\begin{multline} \label{dinsityExVa}
     \left<\hat{A}\right>_t = \\
     \sum_{n,\nu,i} \left<\hat{U}^\dagger\left(t\right)\Tilde{\psi}_{n,\nu,i}\left(t=0\right)\right| \hat{A} \left|\hat{U}\left(t\right)\Tilde{\psi}_{n,\nu,i}\left(t=0\right)\right>
\end{multline}
Equation \ref{dinsityExVa} can be decomposed into the expectation values of the individual components. Each component requires the solution of the time-dependent Schr\"{o}dinger equation for a wavefunction: $ i \frac{d}{dt} \psi = \hat H(t) \psi $, which is then computed by employing short time steps and solving for each step by the Chebyshev approximation \cite{tal1984accurate,kosloff1994propagation,chen1999chebyshev} assuming that $\hat H(t)$ is piecewise constant. The basic evaluation of $\hat H \phi $ for Eq. \ref{eq:Hamiltonian} is carried out using the Fourier grid method for the internuclear coordinate \( r \) \cite{kosloff1983fourier,kosloff1988time} and angular momentum algebra for the rotation. Details of the parameters are given in Table \ref{table-1}.\par
The propagation is therefore decomposed into independent wavefunctions. The initial state for a wavefunction representation becomes:
\begin{multline}
    \ket{\Psi_i\left( r,\theta,\phi;t\right)} = \sqrt{\xi\left(j_i\right)}\sum_{n,\zeta} \sqrt{a_{\zeta,n,\zeta,n}\left(t\right)} \\
    \ket{\Phi_{n,\zeta} \left(r;t\right)}\otimes\ket{n}\otimes \ket{\zeta} \label{eq:endPulse}
\end{multline}

\subsection{Absorbing boundary condition}
In a typical photodissociation experiment, the fragments are detected far from the dissociation point. At the detection point, the particles are characterized by their momentum, which is detected by velocity map imaging \cite{eppink1997velocityMap,shahi2019ultrafast}. Comparable information can be calculated by the asymptotic momentum, amplitude, and direction, at far internuclear distance. We can employ the fact that the position information is not detected and restrict the computation grid in $r$ up to the region where the potential becomes flat. To extract the direction, the phase information must be conserved, since the wavefunction is encoded in angular momentum components and not in direction.

To achieve these goals, we constructed an auxiliary grid in $r$ for each wavefunction component. This grid is flat with potential energy matching the asymptotic potential of the primary grid. The two grids overlap in the asymptotic flat part of the primary grid. At constant time intervals $\tau$, a portion of the asymptotic part of the wavefunction is moved to the auxiliary grid. This is achieved by a smooth transfer function which eliminates spurious reflections. This decomposition is possible due to the linearity of the wavefunction \( \Psi = \Psi_{primary} + \Psi_{aux} \). Both wavefunctions are propagated simultaneously. The auxiliary wavefunction is stored in momentum space where the time propagation is just a phase shift by \( \hat U(t)=\exp( -\frac{i}{\hbar}\frac{\hat P^2}{2m} t ) \).\par
At the end of the propagation, all of the dissociated portion of the wavefunction is stored in the auxiliary grid:
\begin{equation}
\label{psiTfinal}
    \Psi_i\left(r,\theta,\phi;t=t_{final}\right) = \sum _{n_{\tau}} \sum_{\eta,\zeta} \ket{\eta}b^{i}_{\zeta,\eta} \left(r;n_{\tau}\right) \ket{\zeta}
\end{equation}
where \( n_{ \tau} \) is the time index for the wavefunctions that are transferred at different times, \(\eta\) denotes the auxiliary surface (corresponding to the coupled surface $n$), and we composed the coefficients and the dependency of \(r\). More information describe at the appendix section \ref{Abs-App}.

All of the observables are calculated based on Eq.\ref{dinsityExVa}, which can be decomposed to individual wavefunction propagations, leading to the final state, Eq.\ref{psiTfinal}. The thermal average observables are obtained by summation over the individual thermal components calculated for different initial states. 
\begin{equation}
    \left<A\right> = \sum_i \left<A\right>_i 
\end{equation}
When more than one excited state potential is included, each channel is calculated separately.
\vspace{5mm}

The next sections contain the results of several calculations, for description for each observable see appendix section \ref{Obser-App}. The model was set with the parameters from Table \ref{Tal.Para}.  

\begin{table}[t]
\caption{Model parameteres}\label{Tal.Para}
\label{table-1}
\begin{tabular}{|c|c|c|}
\hline 
\multicolumn{2}{|c|}{parameter} & value\tabularnewline
\hline 
\hline 
\multicolumn{2}{|c|}{$\mu$-reduce mass of $F_{2}^{-}$} & $9.5\,amu$\tabularnewline
\hline 
\multicolumn{2}{|c|}{Bond length} & $6.8\,Angstrom$\tabularnewline
\hline 
\multicolumn{2}{|c|}{Rotational Constant} & $8.88\,cm^{-1}$\tabularnewline
\hline 
\multicolumn{2}{|c|}{$\Delta r$} & $0.0464\,bohr$\tabularnewline
\hline 
\multicolumn{2}{|c|}{$\Delta t$} & $5\cdot10^{-2}\,fsec$\tabularnewline
\hline 
\multicolumn{2}{|c|}{Pulse width } & $30\,fsec$\tabularnewline
\hline 
\multicolumn{2}{|c|}{Central wavelength} & $320-390\,nm$\tabularnewline
\hline 
\multicolumn{2}{|c|}{Intensity} & $10^{11}-10^{16}\,\frac{W}{cm^{2}}$\tabularnewline
\hline 
\multicolumn{2}{|c|}{Electronic potentials} & private communication\tabularnewline
\hline 
\multicolumn{2}{|c|}{Temperature} & $20\,K$\tabularnewline
\hline 
\end{tabular}
\end{table}

%%%%%%%%%

\section{Theoretical analysis of experimental observables}

The model is designed to provide a full description of a photodissociation  experiment in a strong laser field. The system is initiated in a thermal state. The initial state resides on the ground bounded electronic and vibronic states with a thermal distribution of  rotational states. The external pulse field couples the rotational states in the ground electronic to rotational states in the excited un-bounded electronic states. We systematically increased the radiation intensity to gain insight on strong field effects. \par
In all calculations, the ground electronic state was set as the singlet \(\Sigma\) bound electronic surface. The excited electronic surface was set to either \(\Sigma\) singlet or \(\Pi\) singlet with the same potential form. In all simulations, the electromagnetic-radiation was chosen to be a transform limited Gaussian pulse with a width of \(30 \, fs\). Wavelengths and intensities were varied. 

\subsection{\(\Sigma\) excited electronic surface}\label{SSresults}

\begin{figure}
  \includegraphics[width=90mm]{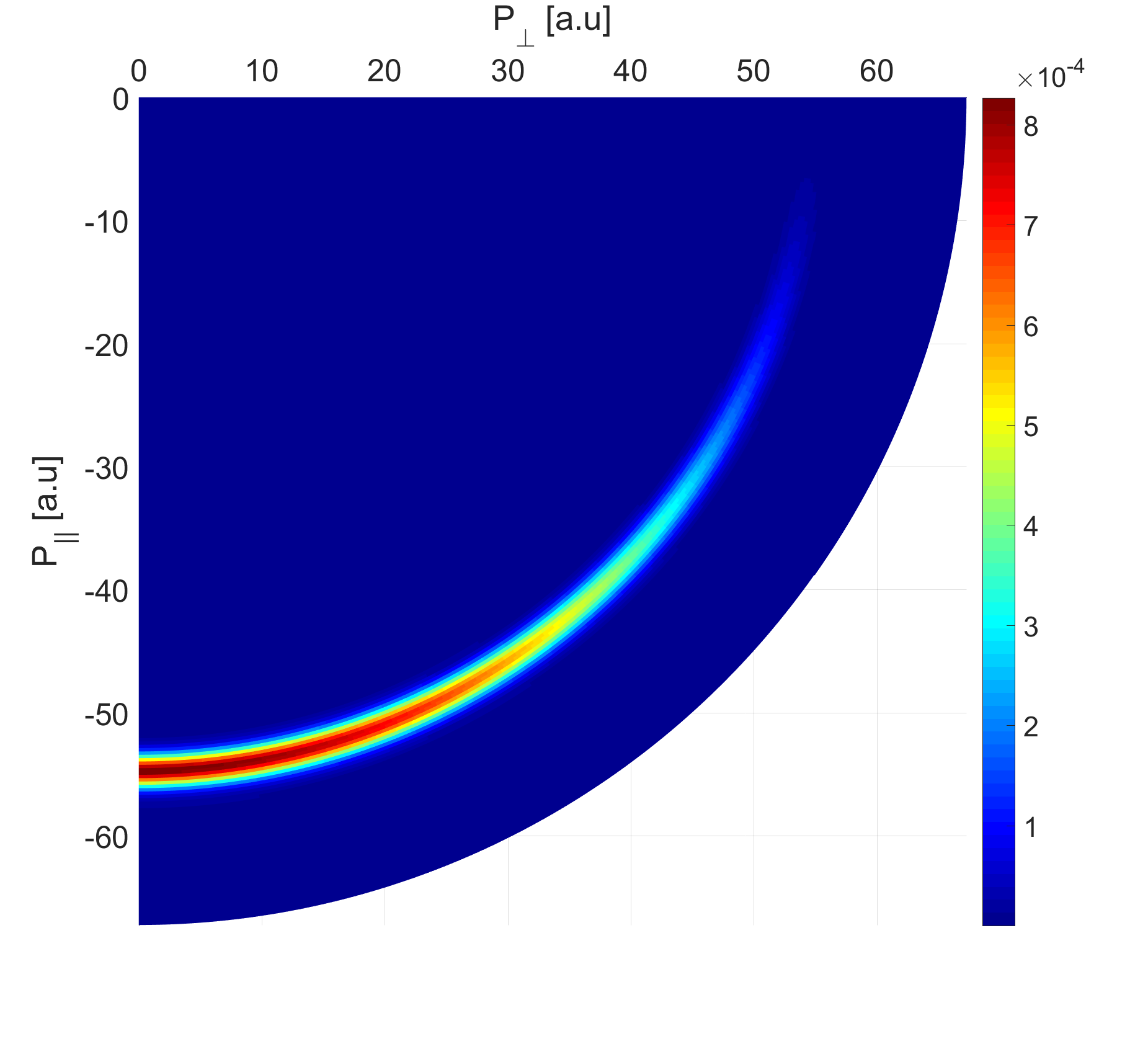}
  \caption{Angular distribution of the outgoing momentum shown as a density plot. The results are for fragments emitted from the transition between two \(\Sigma\) electronic surfaces with low intensity, $I=10^{10}$. The distribution is proportional to $\cos^2\theta$ which indicate one photon process. } \label{SimgaSigmaDist}
\end{figure}

We first examine the dissolution outcome of radiation coupling two \(\Sigma\) electronic surfaces. Figure \ref{SimgaSigmaDist} presents the angular distribution of the fragments' momentum in the parallel and perpendicular directions. For the weak field one-photon transition
in a parallel excitation, the resulting distribution resides mainly in the parallel direction, and is proportional to \(\cos^{2}\left(\theta\right)\), as expected. \par

\begin{figure}
  \centering
  \includegraphics[width=80mm]{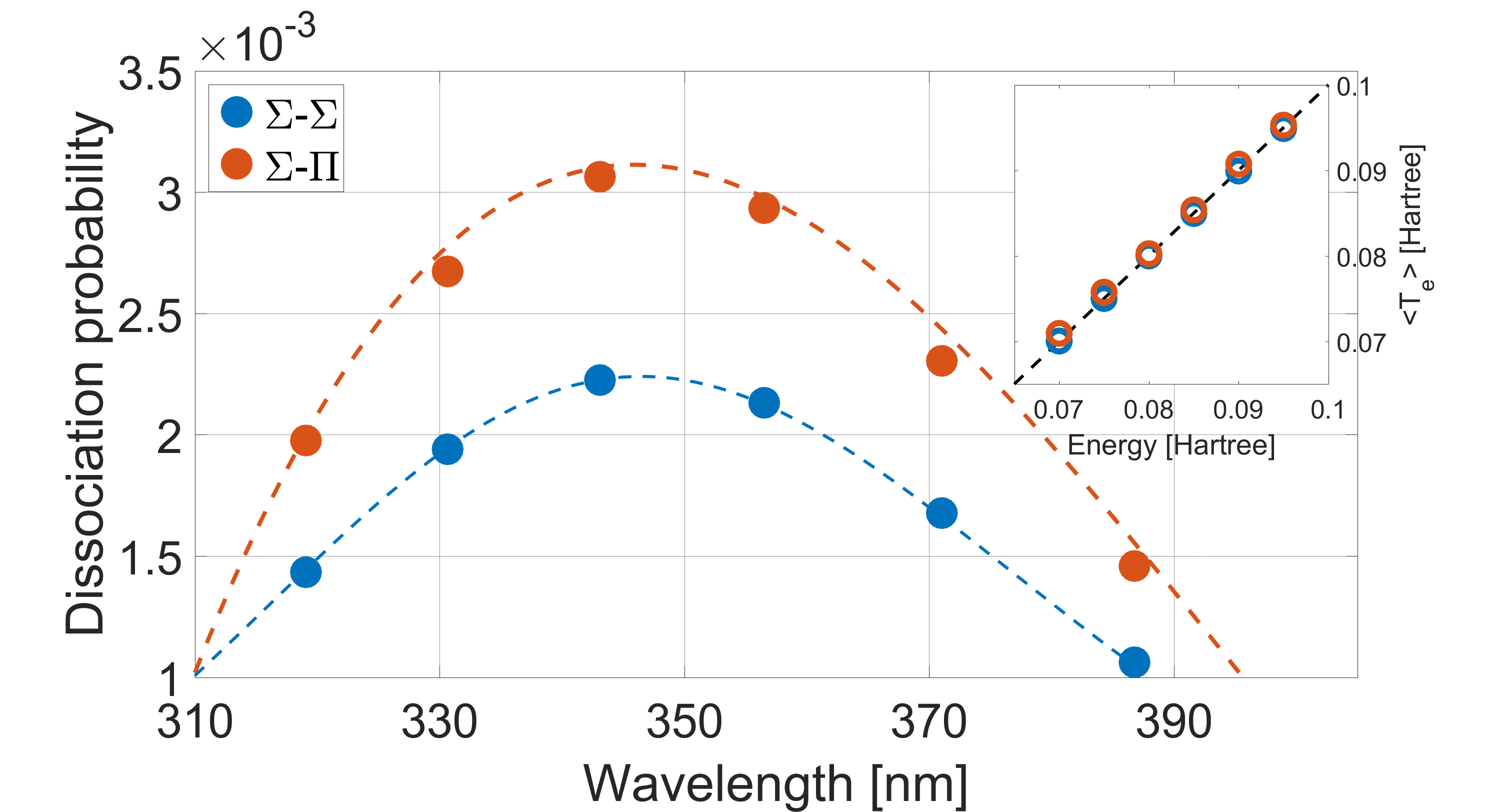}
  \caption{Action spectrum represent the dissociation probability as function of wavelength for \(\Sigma-\Sigma\), blue data, and $\Sigma-\Pi$, red data, electronic transfer. The action spectrum calculated with high intensity in the one photon transition range,$1\cdot 10^{11} \left[W/cm^2\right]$. The inset: Average kinetic energy of the photo-fragments, \(\left<T_e\right>\), as a function of Condon energy. The Condon energy is defined as the vertical energy minus the dissociation energy. } \label{SSKineticEnergy}
\end{figure}

Figure \ref{SSKineticEnergy} presents the dissociation probability as a function of wavelength. Neglecting spontaneous emission, the dissociation probability is equal to the excited state population. The excited state population is also proportional to the absorption energy \cite{GuyPhotodissDynamics,kosloff1992excitation} \(\hbar\omega\Delta N = \Delta E\). By fitting the action spectrum (dashed line), we conclude that the peak of the transition is at \(346 \, nm\). \par

The coupling between the ground state and the excited state depends on the laser wavelength, which determines the Condon point. The Condon point is defined as the radius in which the two coupled potentials cross for a given vertical resonance energy. The vertical resonance energy value is the Condon energy. The inset in Fig.\ref{SSKineticEnergy} displays the correlation between the fragments' average kinetic energy at different wavelengths and the corresponding Condon energy.  \par

Nonlinear multi-photon processes develop when the laser intensity is increased. To explore these processes, we carried out a series of calculations with increasing pulse intensity. The simulations were conducted with the same wavelength, close to the peak of absorption. \par

\begin{figure}
  \includegraphics[width=80mm]{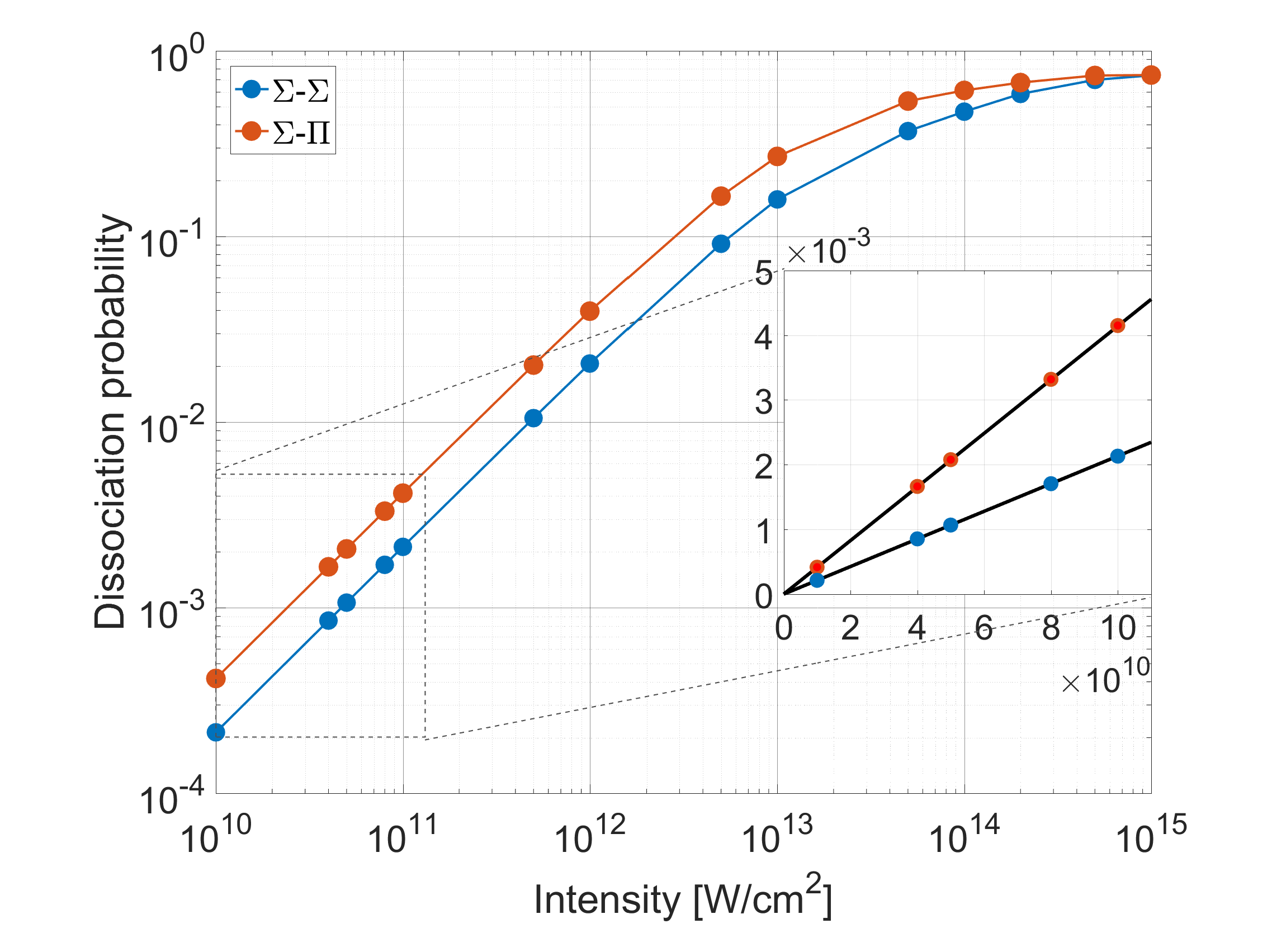}
  \caption{Dissociation probability as a function of laser intensity for $\Sigma-\Sigma$ and $\Sigma-\Pi$ transition in blue and in red respectively. The intensity and the dissociation probability are presented in logarithmic scale. Inset: the linear range, linear fit to the data at black line. } \label{DissProbFigVsInt}
\end{figure}

Figure \ref{DissProbFigVsInt} presents the dissociation probability as a function of laser intensity. The figure includes results of $\Sigma$ and $\Pi$ excited electronic states (see forward). For the case of an  $\Sigma$ excited electronic state, the dissociation probability increases with the laser intensity and saturates. The inset in Fig.\ref{DissProbFigVsInt} presents the dissociation probability for low field intensity. As expected, the dissociation probability is linear with the intensity, fitted in the inset by a black line. Note that a $log$--$log$ scale was used for the main frame, while a linear one was used for the inset.  \par

The number of photons involved in the process reshapes the angular distribution. Each photon transition multiplies the distribution by \(\cos^2\left(\theta\right)\). Therefore, for example at the three-photon transition, excitation--emission--excitation, the angular distribution becomes proportional to \(\cos^6\left(\theta\right)\). Considering the DC component of the distribution, the outcome distribution form three-photon transition has a component of $\cos^4\left(\theta\right)$. 
\par
\begin{figure}
  \includegraphics[width=90mm]{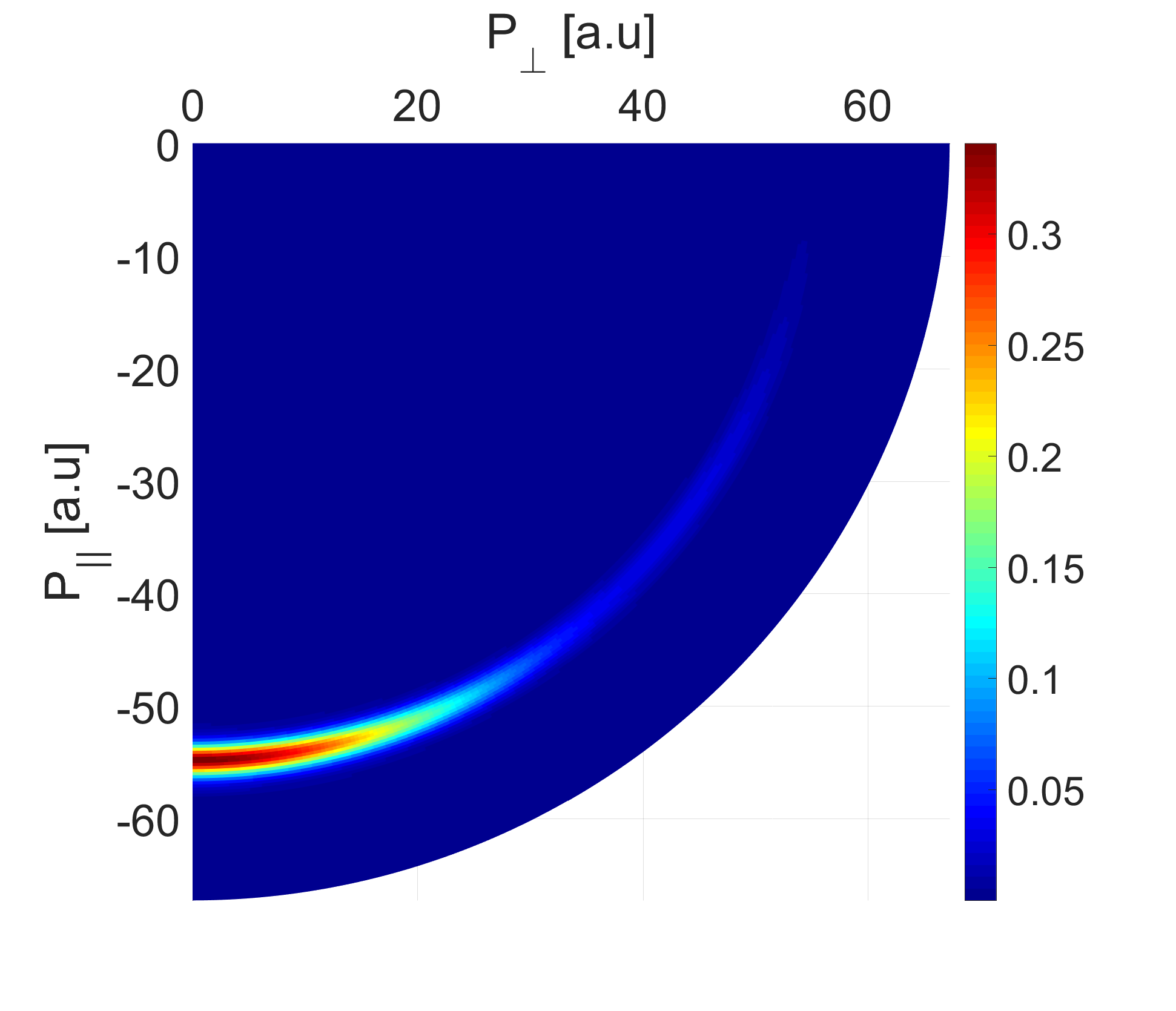}
  \caption{The same as Fig \ref{SimgaSigmaDist} with higher intensity, $I=10^{16} \frac{W}{cm^{2}}$. The distribution is proportional to sum of $\cos^n\theta$ with $n$ larger than $2$. } \label{SSDissProbFigI1e16}
\end{figure}

The full angular distribution can be written as a series of Legendre polynomials  $\mathsf{P}_N\left(\cos{\left(\theta\right)}\right)$, each one caused by different number of photons transition. The series is forgathered at the highest photon transition of the process. Therefore, fitting the angular distribution to a polynomial series indicates the highest multi-photons transition in the process.
\begin{equation}
    p\left(\theta;P\right) = \sum_{{N=2,4,6..}} \alpha_{{N}}\left(P\right) \mathsf{P}_N\left(\cos{\left(\theta\right)}\right)
\end{equation}\label{eqLinearFitSS}
An example of angular distribution at high intensity is presented in Fig.\ref{SSDissProbFigI1e16}. Here, the stronger coupling with the field leads to a concentration of the probability in the parallel direction. This observation fits the classical intuition: when modulation  in a particular direction, the dissociation becomes selective in this direction. 
 
\begin{figure}
  \includegraphics[width=90mm]{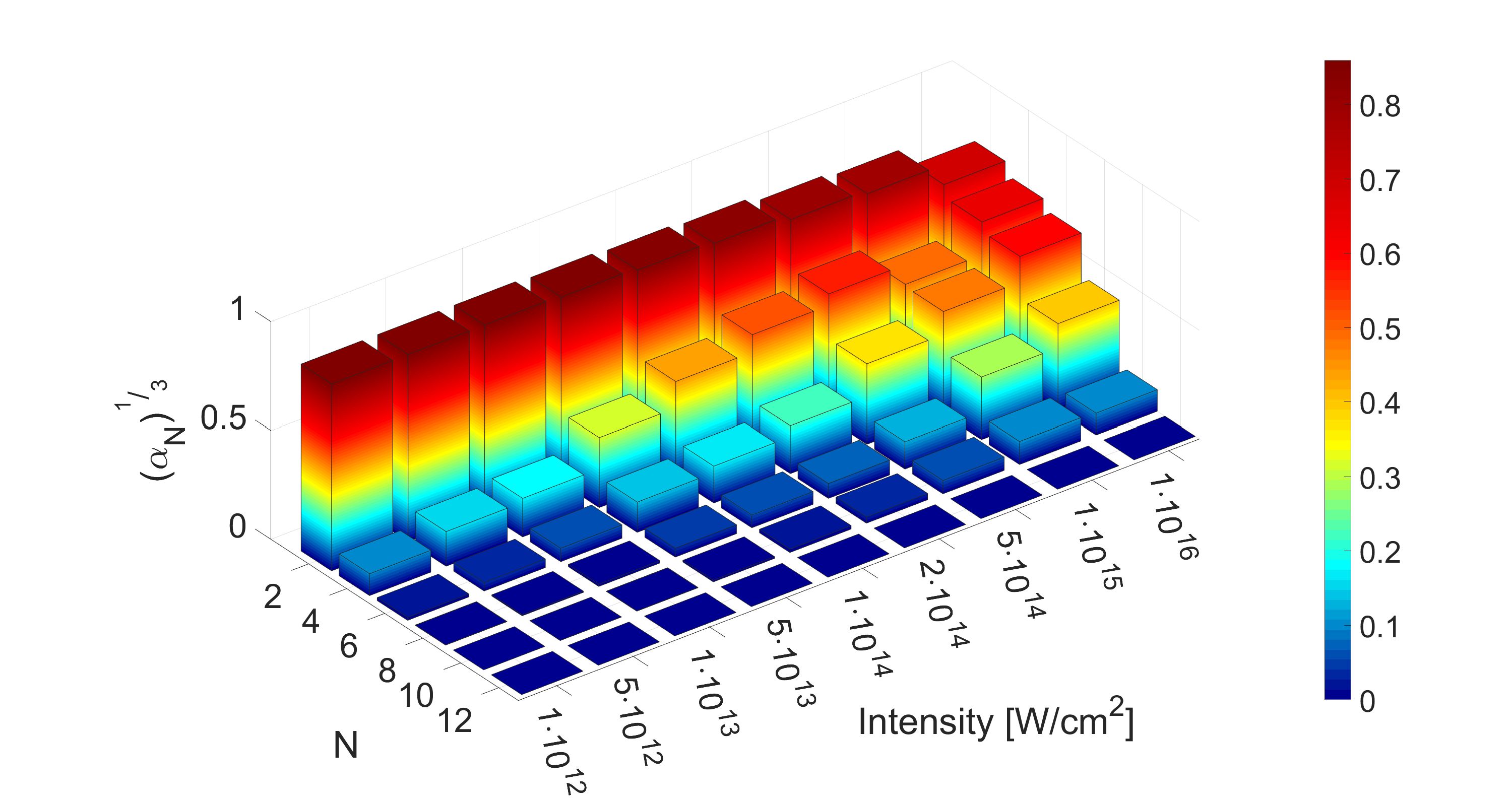}
  \caption{The coefficients of fitting the angular distribution into a polynomial function as function of the intensity and the fitting order. The fitting is done to polynomial function of Legendre polynomials $\mathsf{P}_N\left(\cos{\left(\theta\right)}\right)$. The first expansion coefficient $\alpha_2$ is divided by 3 to make the comparison more visible. Note, the coefficients display after cube root for convenience only. } \label{SSFittingFig}
 
\end{figure}

The difference between the angular distribution at different intensities can be examined by fitting the distribution at the most probable momentum, $P_{mp} (\theta)$. Figure \ref{SSFittingFig} presents the coefficients of \(N=2\) to \(N=10\) for the \(\Sigma\) excited electronic state as a function of laser intensity. The fitting was done using a Fourier transform to expand the angular distribution to $2 \pi$.  

%% SK - The part below seems to me to be less informative and useful. We don't really see multiphoton here - and therefore there is no motion for <E_k>... indeed it is on the 4 digit here...

%%from here:

\begin{figure}
  \includegraphics[width=90mm]{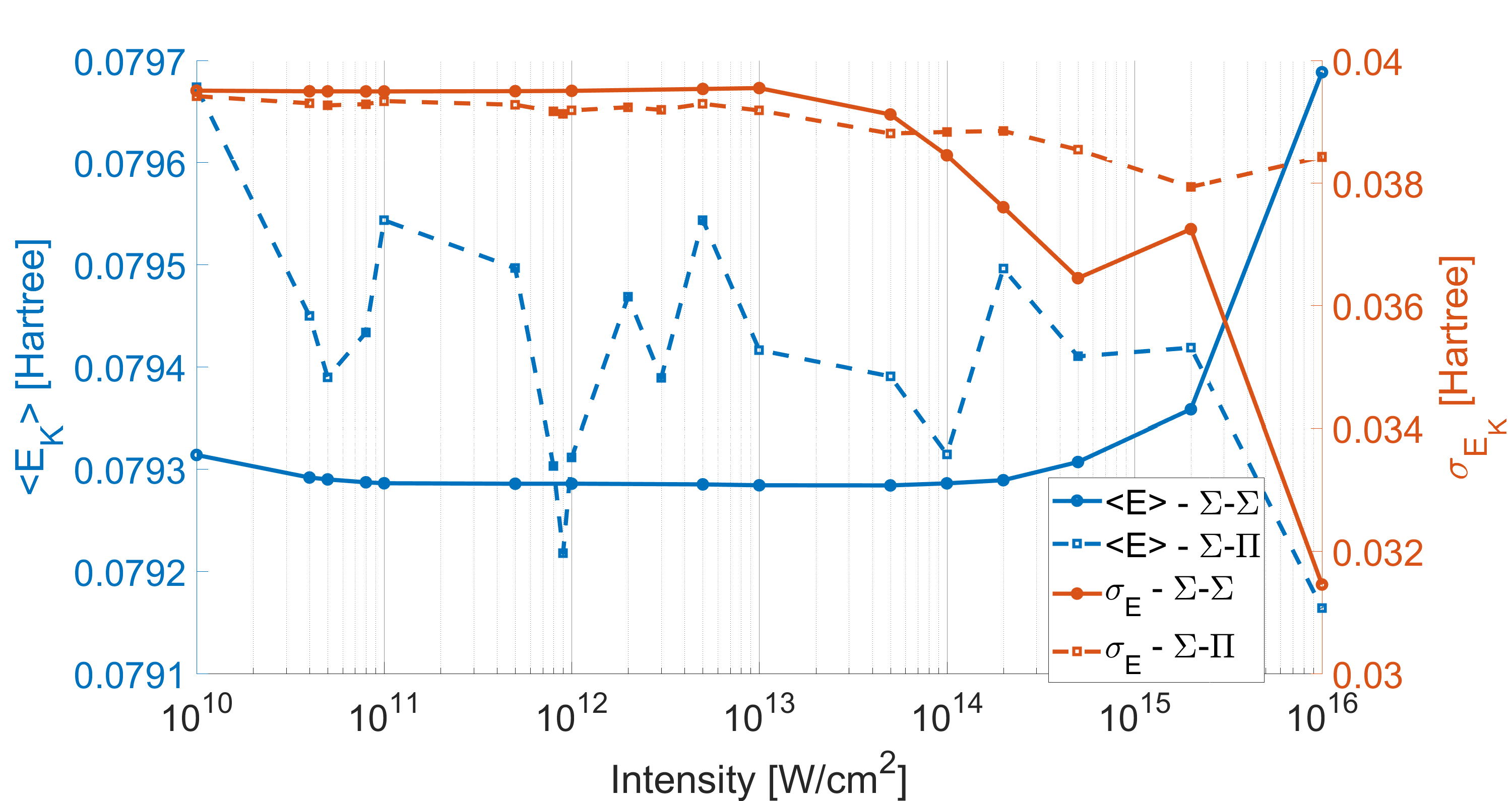}
  \caption{The average kinetic energy release, solid line, and its variance distribution, dashed line. The figure separate to two Y axis, left for the average kinetic energy and the right for the variance distribution. The figure include data of $\Sigma-\Sigma$ transition, in blue, and of $\Sigma-\Pi$ transition, in red. } \label{KEnegry}
 
\end{figure}

Increasing the intensity has a small effect on the kinetic energy of the fragments. Figure \ref{KEnegry} presents the mean value and the variance of the kinetic energy release (KER) for different laser intensities. The KER, for the transition between two $\Sigma$ states (solid blue line), initially decreases with the intensity. This could be a result of the photon-locking process. At very high intensity, we observe an increase in the KER accompanied by a narrow distribution. The order of change in the KER is from low intensity to $1\cdot 10^{15} \left[W/cm^2\right]$ is $\sim 0.1\%$. 

It is essential to note that the extremely high intensities shown are beyond the validity
of the Born--Oppenheimer model employed. 

%%to here. Maybe add this:
%The average kinetic energy release and its distribution are interesting parameters. We calculate these parameters and saw that the intensity does not affect them. Therefore, we conclude that the additional EM field energy invested in changing the rotational energy. 

%%%%%%%%%%%%%%%%
\subsection{\(\Pi\) excited electronic surface}

The quantum number \(\Omega\) changes during the $\Sigma \to \Pi$ transitions. As explained in the previous section, at the \(\Sigma\) electronic surface \(\Omega =0\), whereas  at the \(\Pi\) electronic surface, \(\Omega=\pm1\). Since each excited state level is now doubly degenerate, and the \(Q\,branch\) is allowed, each rotational state in the ground electronic surface is coupled to six states in the excited surface. In this case, the angular distribution of the fragments is expected to be proportional to \(\sin^{2}\left(\theta\right)\) for one-photon transitions.  \par

\begin{figure}
  \includegraphics[width=90mm]{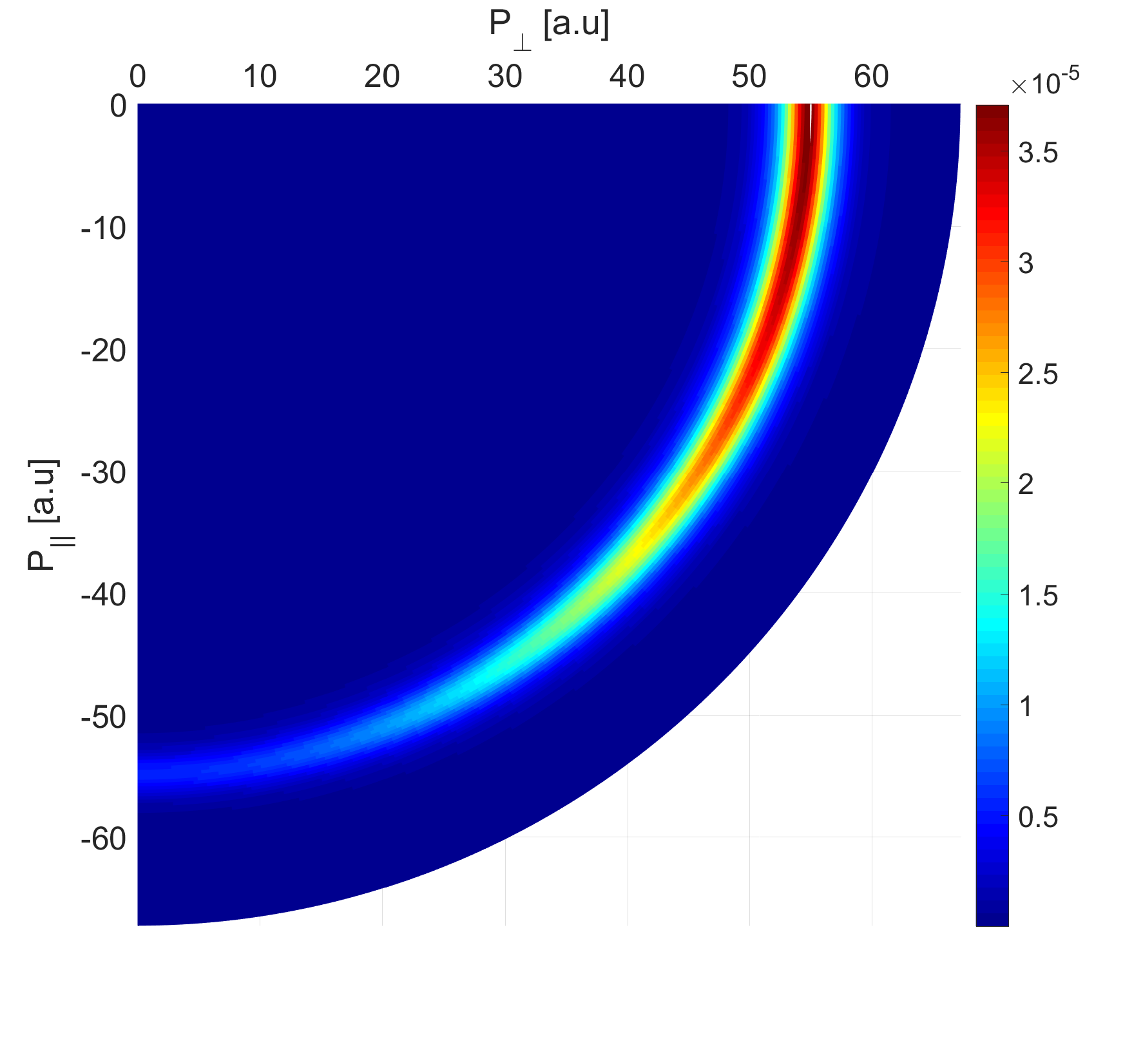}
  \caption{Same as Fig \ref{SimgaSigmaDist} for the transition between ground \(\Sigma\) electronic surface end excited \(\Pi\) electronic surface.} \label{SimgaPiDist}
\end{figure}

Figure \ref{SimgaPiDist} presents the angular distribution of the fragments. A comparison to Fig.\ref{SimgaSigmaDist} reveals that the center of the momentum distribution, and thus the average kinetic energy, is similar for the two symmetries. 

The dissociation probability is influenced by the change of in the excited electronic surface to \(\Pi\). The dissociation probability as a function of the intensity is presented in Fig.\ref{DissProbFigVsInt}. The slope at low intensities, the inset in Fig.\ref{DissProbFigVsInt}, is larger compared to the previous case of on \(\Sigma\) electronic surface. The reason is the larger number of excited states that are coupled to each ground state. Furthermore, this change impacts the saturation intensity of the dissociation probability.     \par

\begin{figure}
  \includegraphics[width=90mm]{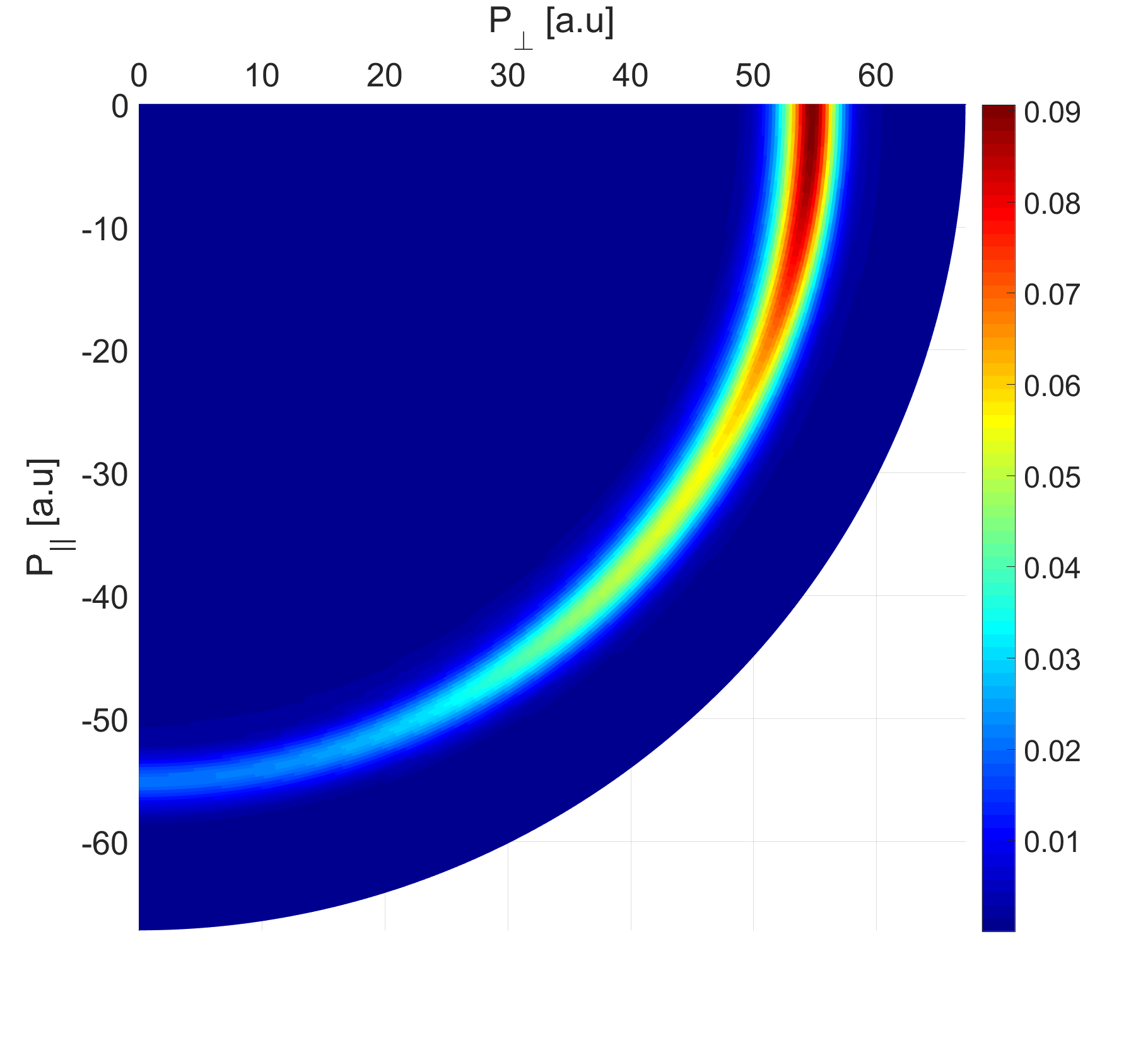}
  \caption{The same as Fig \ref{SimgaPiDist} with higher intensity, $I=10^{16} \frac{W}{cm^{2}}$. The distribution proportional to sum of $\sin^n\theta$ with $n$ larger than $2$. } \label{SPDissProbFigI1e16}
\end{figure}
In the  \(\Sigma-\Pi\) transition, each absorbed photon multiplies the distribution by \(\sin^2\left(\theta\right)\). This can be compared to the \(\Sigma-\Sigma\) case where each absorbed photon multiplies the distribution by \(\cos^2\left(\theta\right)\). It is customary to fit the angular distribution to $I(\theta)=1+\beta P_2(cos(\theta))$. For single-photon excitation, this is also the appropriate form for $\Sigma \rightarrow \Pi$ transition. For multi-photon transition this form cannot fit the results. Therefore, the angular distribution was fitted to a polynomial series in $\mathsf{P}_N\left(\sin{\left(\theta\right)}\right)$, which for even $N$ can be written also as $1-\mathsf{P}_N\left(\cos{\left(\theta\right)}\right)$. Figure \ref{SPDissProbFigI1e16} presents the results for coupling between the \(\Sigma\) and \(\Pi\) electronic states with high-intensity laser.  

\begin{equation}
    p\left(\theta;P\right) = \sum_{N=2,4,6..}\alpha_N\left(P\right) \mathsf{P}_N\left(\sin{\left(\theta\right)}\right)
\end{equation}

\begin{figure}
  \includegraphics[width=90mm]{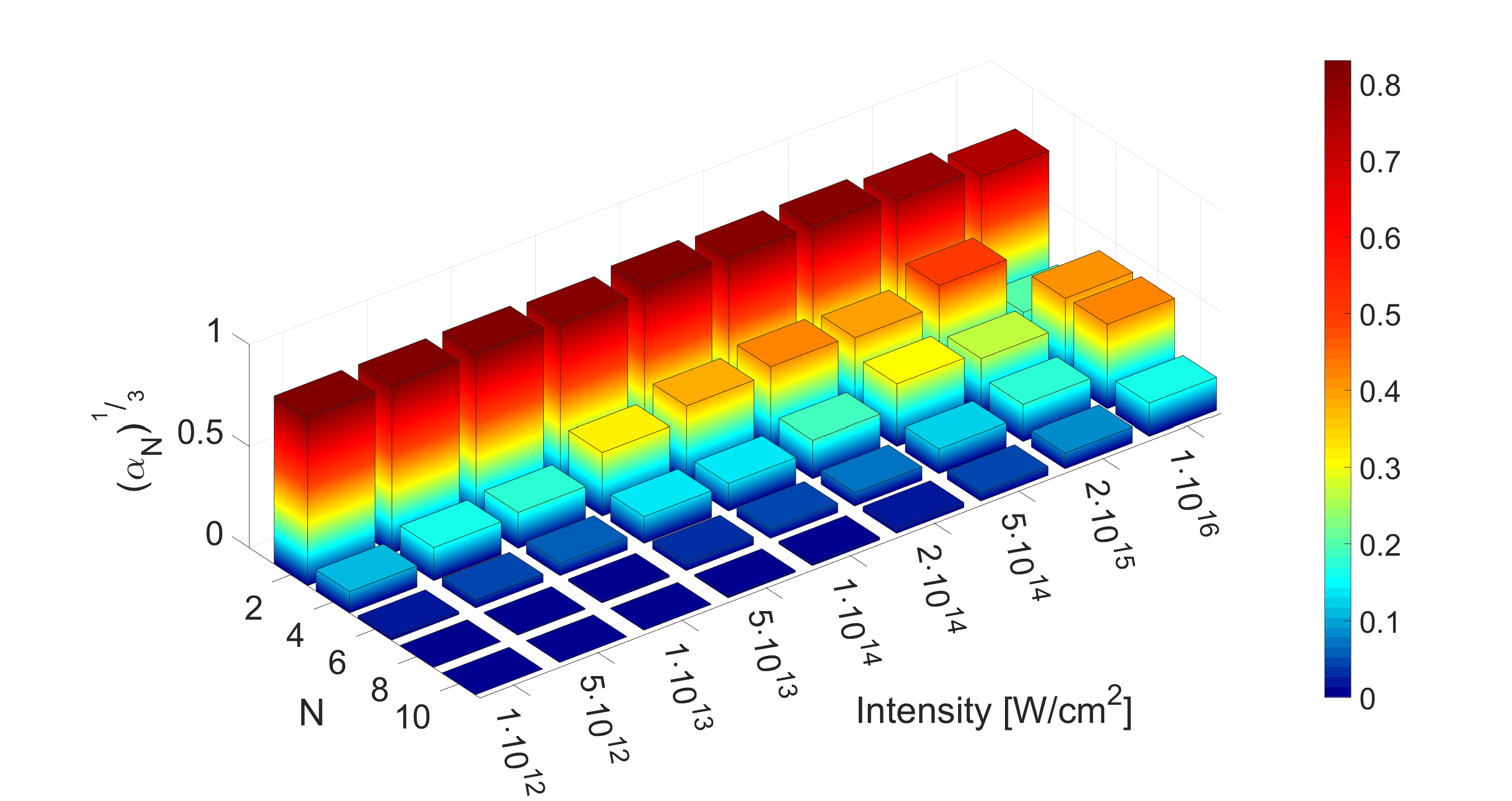}
  \caption{The same as Fig \ref{SSFittingFig} at $\Pi$ excited electronic state. The fitting done to $\mathsf{P}_N\left(\sin{\left(\theta\right)}\right)$. The coefficients present as function of the intensity and the fitting order. Note, the coefficients display after cube root for convenience only.} \label{SPFittingFig}
\end{figure}

Figure \ref{SPFittingFig} presents the coefficients of $N=2$ to $N=10$ for the \(\Pi\) excited electronic state as a function of the laser intensity and the fitting order. By comparing this result to those of the $\Sigma$ excited state, Fig.\ref{SSFittingFig}, we conclude that the multi-photon transition is achieved at a higher intensity for the \(\Pi\) excited state. This difference is a result of the larger effective Hilbert space for the \(\Pi\) excited state.\par 

The KER observations for the transition between $\Sigma$ and $\Pi$ states are presented in Fig.\ref{KEnegry} (dashed blue line). The KER shows non monotonic variations compered to the smooth dependence on intensity for the two $\Sigma$ states. In addition, the overall variation is larger, $\sim 0.6\%$ but still quite small. This behavior could be a result of interference between the excitation pathways in a strong field.

\section{Discussion and conclusion}

The presented model for the photo-dissociation process is based on the exact solution of the time-dependent Schr\"{o}dinger equation. Our survey of existing models of photodissociation showed that they almost always employ very limiting simplifying assumptions. In this paper, we present a new model that aims to describe various significant systems, comparing different electronic state symmetries. The model contains a description of the nuclear rotational motion, and the coupling elements between  rotational states.  

Using this model, we presented the dissociation outcome for two processes, the transition between the $\Sigma$ singlet electronic state to the $\Sigma$ or $\Pi$ singlet electronic state. For each case, we simulated the momentum distribution. 
For weak-field one-photon transitions, we observed a momentum distribution
proportional to $cos^2\left(\theta\right)$ for the $\Sigma-\Sigma$ transition
and $sin^2\left(\theta\right)$ for the $\Pi-\Sigma$ transition.
This is in accordance with 
previous models\cite{zare1972photoejection,seideman1996analysisSeideman,mckenna2012controllingH2} based on perturbation theory. 

Our new model simulate high-laser-intensity processes that lead to multi-photon transitions. We showed that the main effect of increasing the number of photons involved in the process reshapes the angular distribution. For parallel transitions, the angular distribution becomes sharper, indicating a more aligned distribution. In addition, the average KER and its distribution are only weakly dependent on the laser intensity. 

In conclusion, we constructed a new model that can describe a wide range of processes. We advocate the view that the key to understanding a complex dynamical system is to rely on experience gained from small and simple models. Therefore, the results here are presented for two transition cases, between two $\Sigma$ states and between $\Sigma$ and $\Pi$ states. Consistent with prior knowledge, the main difference between the two cases is the angular distribution symmetry. Furthermore, our new model shows that the threshold on the intensity were multi photon transition are occur is lower in the $\Sigma$ excited state then in the $\Pi$ case.

Our model opens the way to exploring the dissociation process of systems with high complexity. For simplicity, the current model introduces singlets as its electronic states. In future studies, we will describe the system using doublet states, which are more suitable to the \(F_2^-\) case. Moreover, spin effects should be taken into account, such as spin--orbit interactions. One might also notice that the model takes advantage of a lower temperature to reduce the size of the required Hilbert space. Higher temperature enlarges the number of potentially populated rotational states, resulting in a larger Hilbert space. Such a simulation will require modified methods such of the use of random phase wavefunction\cite{kallush2015orientation}. 
Furthermore, using coherent control can lead to new and fundamental results\cite{mcdonald2016photodissociation}.  
\par

\begin{acknowledgments}
Daniel Strasser, Robert Moszynski, 
Iwona Majewska, Aviv Aroch, This research was supported by the Israel Science Foundation (Grant No. 510/17).
\end{acknowledgments}
\section{Appendix}

\subsection{Absorbing boundary conditions - Appendix}\label{Abs-App}

\begin{figure}
  \includegraphics[width=80mm]{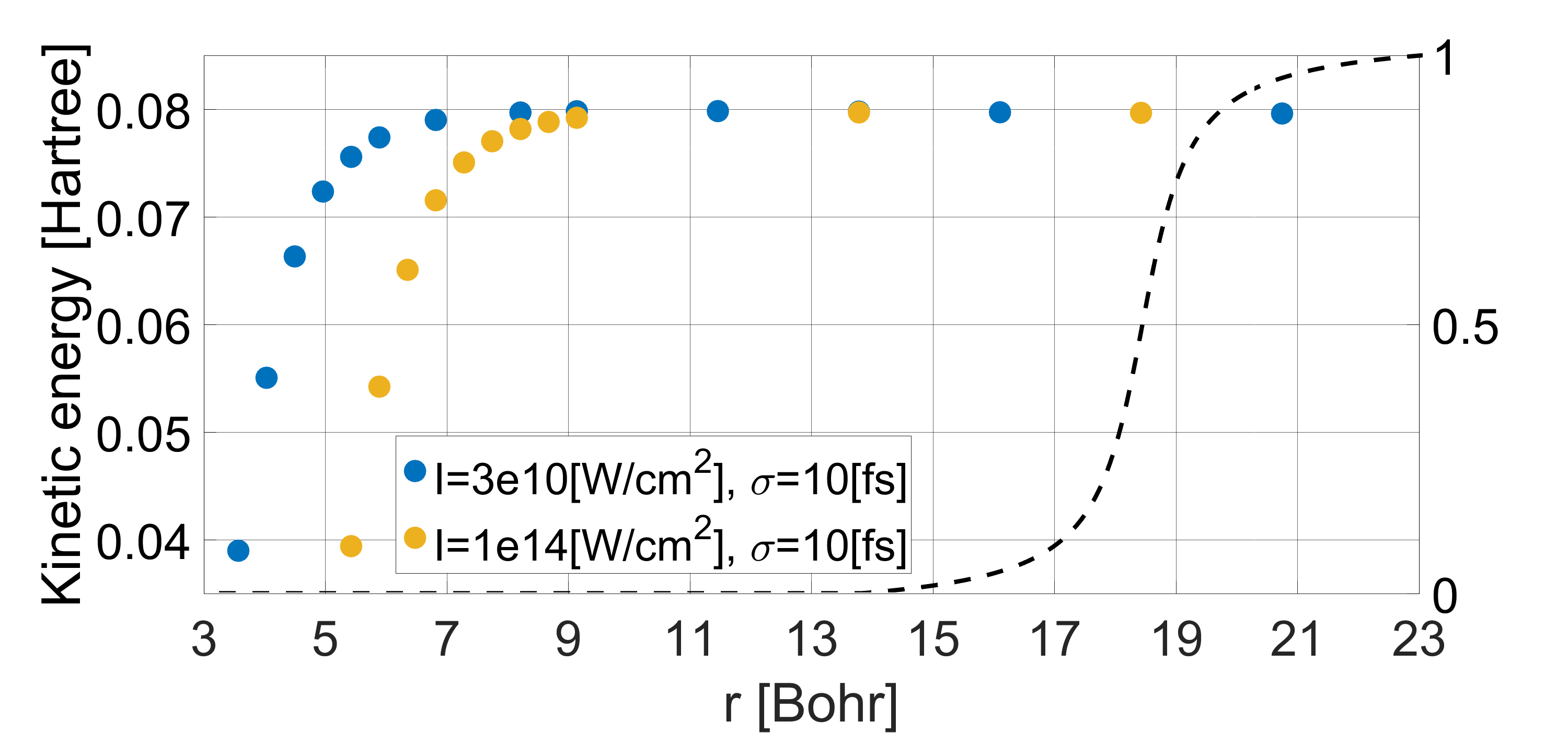}
  \caption{Left axis: Final average kinetic energy calculated for different final states as a function of the starting position of the transfer window. Right axis: Represents the  transfer operator between the grid $n$ to the respectively auxiliary grid.}
  \label{fig:psiFlat}
\end{figure}

To minimize computation effort the computation grid has to be cut at a predetermined
radius. Since we want to analyze the outgoing asymptotic data, we cannot use the common option of absorbing the wavefunction at the end of the grid 
\cite{kosloff1986absorbing,muga2004complex}. 
As an alternative we employ an auxiliary grid which overlaps the primary grid which we term transition window. The position of this window is optimized to minimize the primary grid without damaging the momentum distribution.
The average kinetic energy is then calculated for different window radius points. 
Figure \ref{fig:psiFlat} presents the average kinetic energy as a function 
of the window's starting radius. 
The propagation of the state at the auxiliary surface is a free propagation; hence the momentum distribution will not change with time for  large \(r\). Therefore, the part that absorbs at each time can be restored independently.\par
The total dynamical calculation was stopped when at least $99\%$ of the excited state population  moved to the auxiliary surface. 
%The transfer step is operated only when the population inside the window is not neglected to minimize numeric errors.\par 

\subsection{Theoretical analysis - observables}\label{Obser-App}

For each excited electronic surface we calculated the dissociation probability on the corresponding virtual surface \(\eta\). \par
The accumulated probability for initial state \(i\) 
\begin{multline*}
    \mathcal{P}_{i,\eta} = \\ \left<\Psi_{i,\eta}\left(r,\theta,\phi;t=t_{final}\right)| 
     \Psi_{i,\eta}\left(r,\theta,\phi;t=t_{final}\right)\right> =
\end{multline*}
\begin{equation}
      =\int_0^\infty dr\sum_{m,\Omega,\tau} \left| \sum_j {b}^{i}_{\zeta,\eta} \left(r;\tau\right)    \right|^2 \label{DissProbEq}
\end{equation}

\(\mathcal{P}_{i,\eta}\) is a joint probability starting in initial state \(i\) and dissociation at electronic state \(\eta\). Note that the wavefunction is normalized with Boltzmann coefficients. The first summation, over the quantum number \(j\), is due to the coherence between different components of a given initial thermal state. \par

Asymptotically the total dissociation probability is equal to the missing probability at the ground state, this was been verified for consistently. \par
The momentum angular distribution is proportional to the experimental velocity distribution. The calculation requires to change the representation from the radial degree of freedom to the corresponding momentum, the change is done by using Fourier transform. Additionally, each rotational state is explicitly written in the angular variables.
\begin{multline}\label{psi_endT}
    \Tilde{\psi}_i\left(p,\theta,\phi;t=t_{final}\right) = \\ \sum_{\eta,\zeta}\sum _\tau \Tilde{b}^{i}_{\zeta,\eta} \left(p;\tau\right) \cdot \left|\eta\right> \cdot D^{j}_{m,\Omega}\left(\theta,\phi\right)  
\end{multline}
It is important to notice that in the momentum representation, there is a relative phase between different virtual time parts, \(\tau\).\par
The distribution of each virtual surface and initial state is calculated as follows:  
\begin{equation}
   \mathcal{D}_{i,\eta} \left( p,\theta,\phi \right) = \sum_{m,\Omega} \left| \sum_j \left(\sum_\tau \Tilde{b}^{i}_{\zeta,\eta} \left(p;\tau\right)   \right)\cdot  D^{j}_{m,\Omega}\left(\theta,\phi\right)  \right|^2
\end{equation}
The summation over \(j\) is a quantum summation and conserve phases, due to the coherence between different values that generate during the dynamic. In contrary, the summation over \(m\) and \(\Omega\) are classical since there are no physical coherences between different values. 
\begin{multline}\label{eq:AngularMomentumDistribtion}
   \mathcal{D} \left( p,\theta \right) = \\
   \int d\phi \sum_{i} \sum_{m,\Omega,\eta} \left| \sum_j \left(\sum_\tau \Tilde{b}^{i}_{\zeta,\eta} \left(p;\tau\right)   \right)\cdot  D^{j}_{m,\Omega}\left(\theta,\phi\right)  \right|^2
\end{multline}
We integrate over \(\phi\) since there is no dependency on this angle.\par

\vspace{3mm}

The quantum kinetic energy is calculated as

\begin{equation}
    \mathcal{T}_{} = \int_0^\infty  dp \int_0^{2\pi} d\theta \frac{p^2}{2\mu} \mathcal{D} \left( p,\theta \right)
\end{equation}\label{kineticEq}

\bibliography{mybibliography}

\bibliographystyle{unsrt}

\end{document}